\documentclass[%
 aip,
 jmp,%
 amsmath,amssymb,
 reprint,%
]{revtex4-2}

\usepackage{graphicx,color,mhchem}
\usepackage{dcolumn}
\usepackage{bm}
\usepackage{placeins}
\usepackage[normalem]{ulem}

\begin{document}
\title[]{Ablation of black-Si by (Gauss-)Bessel femtosecond laser beams}

\author{Nan Zheng}%
\author{Hsin-Hui Huang}%
\author{Nguyen Hoai An Le}
\author{Tomas Katkus}%
\author{Haoran Mu}%
\author{Soon Hock Ng}
\author{Thumula Ranaweera} 
\affiliation{Optical Sciences Centre, School
of Science, Swinburne University of Technology, Hawthorn, Victoria 3122, Australia }%
\author{Darius Gailevi\v{c}ius}
\author{Dominyka Stonyt\.{e}$^*$} \email{dominyka.stonyte@ff.vu.lt; corresponding author}
\affiliation{Laser Research Center, Physics Faculty, Vilnius University, Saul\.{e}tekio Ave. 10, 10223 Vilnius, Lithuania}
\author{Saulius Juodkazis$^*$} \email{sjuodkazis@swin.edu.au; corresponding author}
\affiliation{Optical Sciences Centre, School
of Science, Swinburne University of Techology, Hawthorn, Victoria 3122, Australia }
\affiliation{Laser Research Center, Physics Faculty, Vilnius University, Saul\.{e}tekio Ave. 10, 10223 Vilnius, Lithuania}
\affiliation{World Research Hub Initiative (WRHI), Institute of Science Tokyo, School of Materials and Chemical Technology, Tokyo Institute of Technology, 2-12-1, Ookayama, Meguro-ku, Tokyo 152-8550, Japan}

\date{\today}
            
\begin{abstract}
Laser machining and modification of black-Si (b-Si) by femtosecond laser Gaussian (G-) and Gauss-Bessel (GB-) beams are compared at a wavelength of 1030~nm. The GB-beam was generated using a diffractive axicon lens and $10^\times$ de-magnification optics. It was found that modification of b-Si well below (a factor $\sim50^\times$) the single pulse ablation fluence of 0.2~J/cm$^2$ was possible, corresponding to ablation/melting of nano-needles. The width of modification was almost independent of pulse energy/fluence and had a width of $1/e^2$-intensity profile at the melting regime. For the GB-beam, the smallest width of laser modification at $\sim$0.2~J/cm$^2$ threshold (at the center core) was close to the FWHM of the core of the GB-beam. The aspect ratio of the ablated groove on the surface of b-Si made by GB-beam was twice as large --up to 8-- compared to that achievable with G-beam, and it was at lower fluence of $\sim 4$~J/cm$^2$ ($\sim50^\times$ reduction). Reflectivity of two-side nanotextured b-Si on plasma-thinned 70-$\mu$m thick Si was strongly reduced in the near-IR range, reaching transmittance $> 95\%$ at 1.7-2.1~$\mu$m wavelengths.   
\end{abstract}

\keywords{Axicon, black Si, ablation, threshold of ablation}
\maketitle
\tableofcontents
\section{\label{intro}Introduction}
Black-Silicon (b-Si) is a surface-modified form of silicon (Si). It is termed "black" because it appears visually black due to its ability to absorb in the visible wavelength range with the surface containing a random nano-needle pattern~\cite{Huang_2007,16aplp076104,Fan_2020}. The typical method for producing it involves plasma etching of silicon wafers. B-Si was produced using a standard gas-based \ce{SF6}/\ce{O2} etching process,  widely available in the context of standard semiconductor industrial processing~\cite{Eisele1981,Jansen_2001,Nguyen_2020}. The same results can be achieved by using gas breakdown with a femtosecond (fs) laser. However, this method tends to result in an S-doped surface~\cite{Mazur}. Other production methods include electrochemical~\cite{Liu_2014b,Bilyalov2003}, stain etching \cite{Martin-Palma_2001,Saadoun2002}, and metal-assisted chemical etching, which is favored in recent years due to low fabrication costs and versatility in generating high aspect ratio nanostructures. \cite{Um_2015,Huo_2020,Soueiti2023}. One of the applications of b-Si is Si solar cells for the improvement of energy conversion due to the reflectivity of b-Si can reach values below $1\%$ over the visible spectral range~\cite{Halbwax_2008,nguyen2011black,Oh_2012,Otto_2015,20mte100539}, other applications includes photodetectors~\cite{Lv_2018,Zhao_2023a}, sensors~\cite{Tan_2019}, and even biocidal applications of nanotectured surfaces - mechanical antibiotics~\cite {13nc2838}. The height of nano-needles is defined by the etching time, ranging from 250~nm (15 min) to 450~nm (30~min)~\cite{25srm316}.  

In order to understand the energy deposition efficiency on such unusual surfaces, it is important to understand how laser machining behaves on such anti-reflective materials and acquire a clearer comprehension of the ablation, amorphization, and melting thresholds~\cite{Jan}. If laser ablation/cutting can utilize smaller pulse energies, then the propensity of self-focusing can be reduced. The increase in efficient energy deposition is of significant interest, especially for Si. 
For conventional materials, localized melting of the solid may result in volume expansion, which leads to structural failure. However, in very few cases, such as Si, the mass density of molten Si is higher than solid at 2.57~g/cm$^{-3}$ and 2.33~g/cm$^{-3}$, respectively. Other such materials include Ge, Ga, Bi, Sb~\cite{Greenwood_1997}, some compounds, such as \ce{H2O}, or some alloys, e.g., certain Ce-- or Bi-- based alloys~\cite{perkins_geoffrion_biery_1966,Khairulin_2010}. 
Hence, controlled (re-)melting inside the volume of solid Si can be achieved without volume expansion and avoid structural failure. This is promising for hyper-doping of Si by ion implantation and laser annealing~\cite{YANG2017,Li2021,22040012,KAUP}. 
Using beams with a long axial extent of the focal region, such as Bessel-like beams, offers an advantage in efficiently modifying a structured surface, such as b-Si, compared to standard Gaussian beams~\cite{Nowack_2012,Duocastella_2012,Stoian_2018,Rao_2024}. With high aspect ratio and precision, Bessel beams are becoming increasingly popular~\cite{Francois,Dudley} after an early demonstration of 3D capability of fs-laser modifications of transparent materials by Bessel beam~\cite{01jjapL1197,Stoian_2018}. Bessel beams can be generated by several different approaches~\cite{Khonina,18jo085606,Rao_2024}.

Here, we demonstrate the possibility for thinning Si down to tens-of-$\mu$m thickness using dry plasma etching technique and further modify the surface by adding b-Si nano-textures on either one or both surfaces. This texturing further reduces the reflectance of b-Si, achieving near 100\% transmittance for 70-$\mu$m thick double-sided b-Si sample at a wavelength of $\sim 2~\mu$m. We investigated laser structuring by ablation on the anti-reflective b-Si surfaces. It was found that surface modification through melting of the nano-needles occurs below the ablation threshold of Si (0.2~J/cm$^2$~\cite{Bonse,Jan}), extended to fluences by an order of magnitude smaller using Gaussian (G-) as well as Gauss-Bessel (GB-)beams. In order to compare the energy deposition of G- and GB-beams, geometrical analysis of the focusing and ablation sites was carried out. It is shown that the Liu method of ablation threshold determination~\cite{Liu82} is applicable to the central core of the GB-beam.
\begin{figure}[tb]
\centering\includegraphics[width=\textwidth]{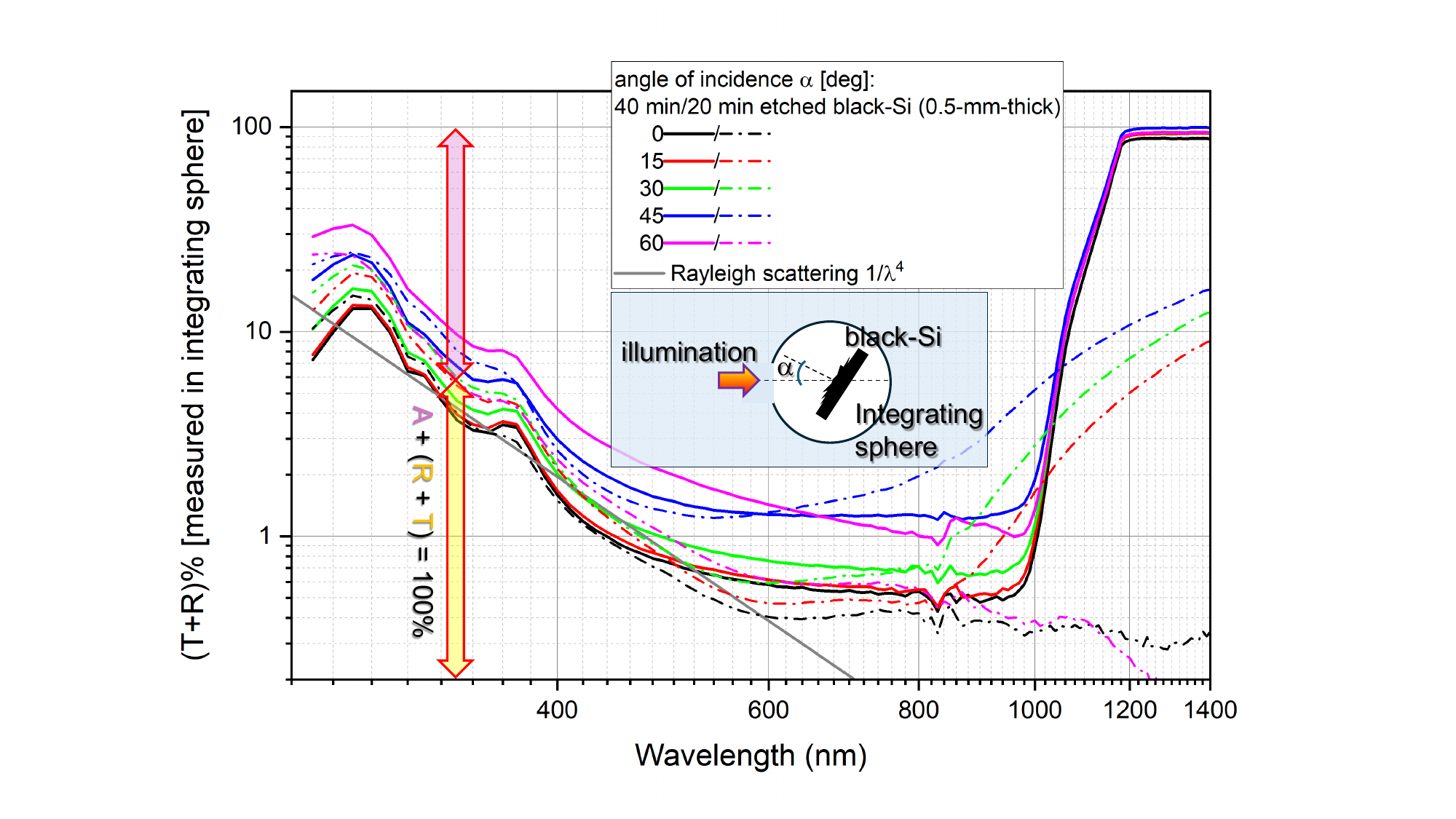}
\caption{\label{f-angle} Absorbance $A = 1- (R +T)$ spectra measured with integrating sphere by detection of reflectance $R$ and transmittance $T$ of b-Si (40 and 20 min etched; one side etched of 0.5-mm-thick Si wafer); measured with UV-Vis spectrometer (Lambda 1050 UV/Vis, PerkinElmer). 
The inset shows schematics of measurements. }
\end{figure}
\begin{figure}[tb]
\centering\includegraphics[width=0.99\textwidth]{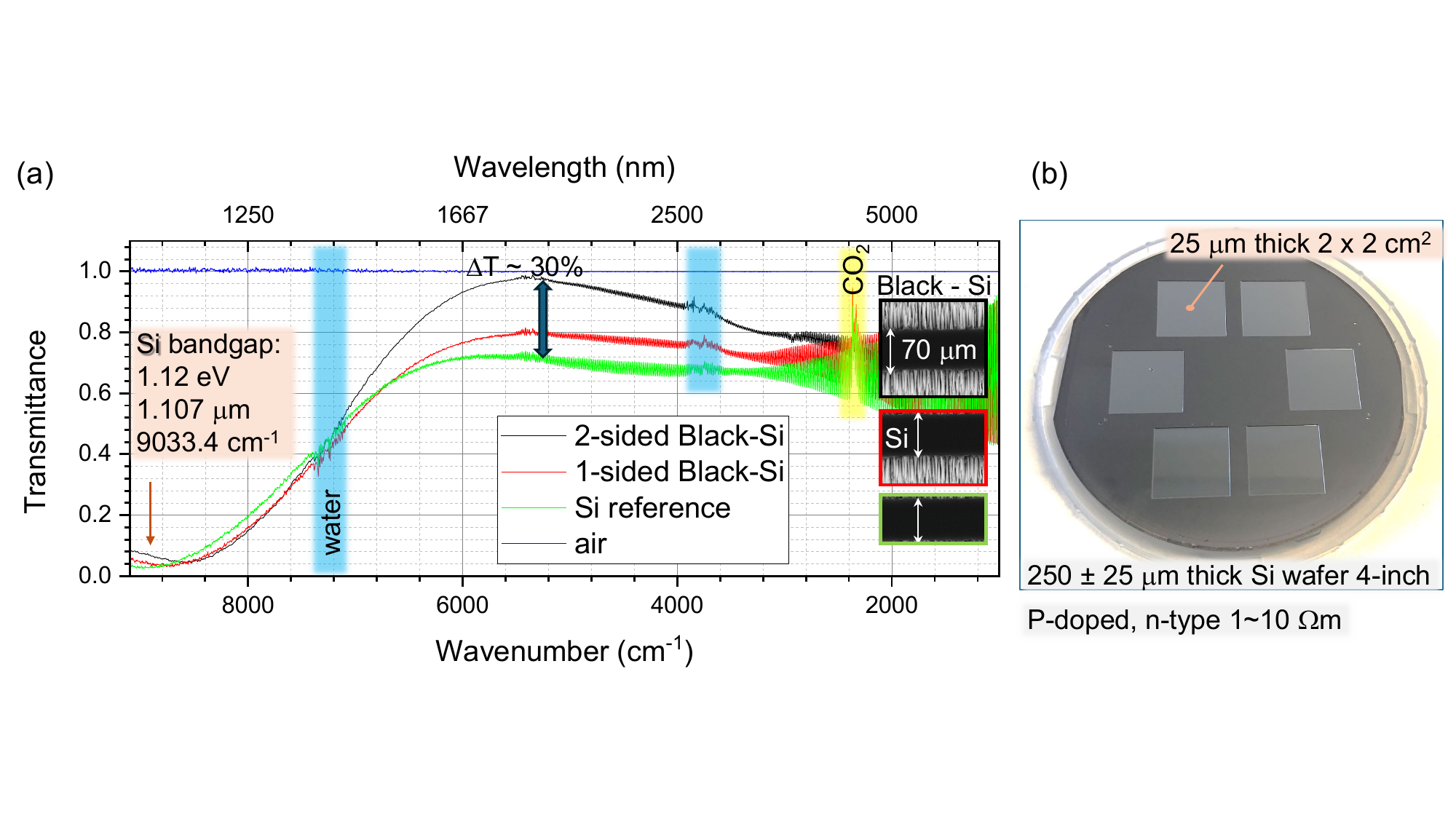}
\caption{\label{f-2side} (a) IR transmittance of one and two-sided b-Si (70-$\mu$m-thick; see schematic thumbnail images as insets). (b) Plasma thinned Si wafer from 250~$\mu$m down to 25~$\mu$m at the selected regions. Direct-write mask was used to define the pattern. Silicon etching was carried out using \ce{SF6} gas in a Plasmalab100 ICP380 reactive ion etching system (Oxford Instruments). Over a 30-minute process, approximately 250~$\mu$m of silicon and 5~$\mu$m of photoresist were etched, corresponding to an estimated silicon etch rate of $\sim$8.3~$\mu$m/min. These values were approximated using a profilometer (Ambios XP-200).
}   
\end{figure}
\begin{figure}[tb]
\centering\includegraphics[width=0.9\textwidth]{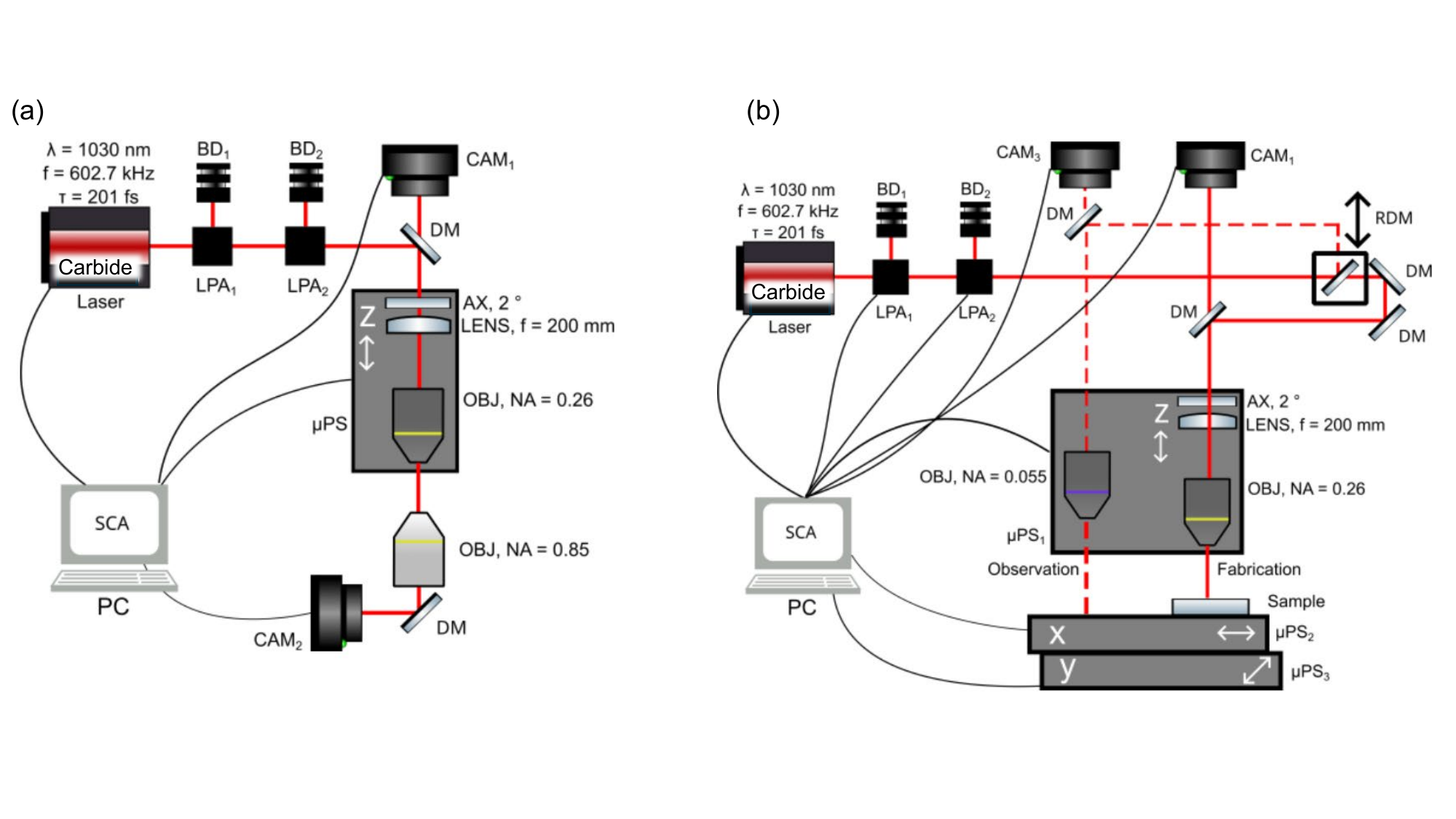}
\caption{\label{f-fab} Laser micro-machining setup (WOP, Lithuania) based on Carbide-40W fs-laser (Light Conversion, Lithuania). (a) Alignment of GB-beam. (b) Fabrication and observation solution; LPA1,2 are laser power attenuators, BD1,2 - beam dumps, CAM1,2,3 – observation CCD cameras, DM –
dichroic mirror, RDM – removable dielectric mirror, AX - diffractive axicon, $\mu$PS1,2,3 – micro-
positioning stages and OBJ are objective lenses with numerical aperture $NA$; the final objective lens has f-number of $f_\# = F/D = 1/(2NA) = 1.92$ with focal length $F = 20$~mm and diameter of entrance pupil $D = 10.4$~mm (10$^\times$ Mitutoyo Plan Apo NIR). The solid red line in (b) indicates the fabrication laser beam path, and the dashed line
is the sample observation beam path. These schematics do not show a mirror that switches beam paths. The observation beam path in (b) is used for G-beam machining.}
\end{figure}

\section{Experimental: samples and methods}

\subsection{Black-Si: plasma thinning and reflectance in visible-to-IR spectral range}

Black-Si (b-Si) is produced by dry plasma etching using a simple \ce{SF6}/\ce{O2} gas mixture~\cite{16aplp076104}. 
Silicon thinning of a standard $300, 500~\mu$m wafer down to tens-of-$\mu$m has become a fast and simple fabrication step. It starts with lithography performed using AZ 4562 photoresist: spin-coated at 1500\,rpm for 30\,s and soft-baked at 110\,\(^\circ\mathrm{C}\) for 4--5\,min. A photomask was generated using the direct-write system (Micropatterning SF100 XPRESS) and aligned with the substrate. UV exposure was conducted using an ABM UV flood light source at 450\,mJ/cm\(^2\). Development was completed in undiluted AZ 726 MIF for 12\,min, with visual inspection to confirm full pattern formation. A post-development bake was ramped from 90\,\(^\circ\mathrm{C}\) to 140\,\(^\circ\mathrm{C}\) at 1\,\(^\circ\mathrm{C}\)/min, held at 140\,\(^\circ\mathrm{C}\) for 5\,min, and cooled gradually. The final resist thickness was approximately 10.9\,$\mu$m. 

Silicon etching was carried out using \ce{SF6} gas in a Plasmalab100 ICP380 RIE system (Oxford Instruments). Over 30 minutes, approximately 250\,$\mu$m of silicon and 5\,$\mu$m of photoresist were etched, corresponding to an estimated etch rate of $\sim$8.3\,$\mu$m/min. These values were approximated using a profilometer (Ambios XP-200), and the process remained stable with no observed plasma flickering or reflected power issues.  

Figure~\ref{f-angle} shows the angular dependence of the absorbance of b-Si with different heights of nano-needles, corresponding to 40~min and 20~min plasma etch. For a broad spectral range of 450-1000~nm, low reflectance $R < 1\%$ (large absorbance $A > 99\%$) was observed even up to angles of incidence $\theta_i\sim 30^\circ$. 
Figure~\ref{f-2side} shows the infrared (IR) transmittance of one-sided and two-sided b-Si samples with a Si wafer thickness of $\sim$70-$\mu$m. The texturing further increased the transmission in the IR region from 80\% to nearly 100\% for the one-sided and two-sided b-Si structures, respectively.

\begin{figure}[tb]
\centering\includegraphics[width=0.85\textwidth]{
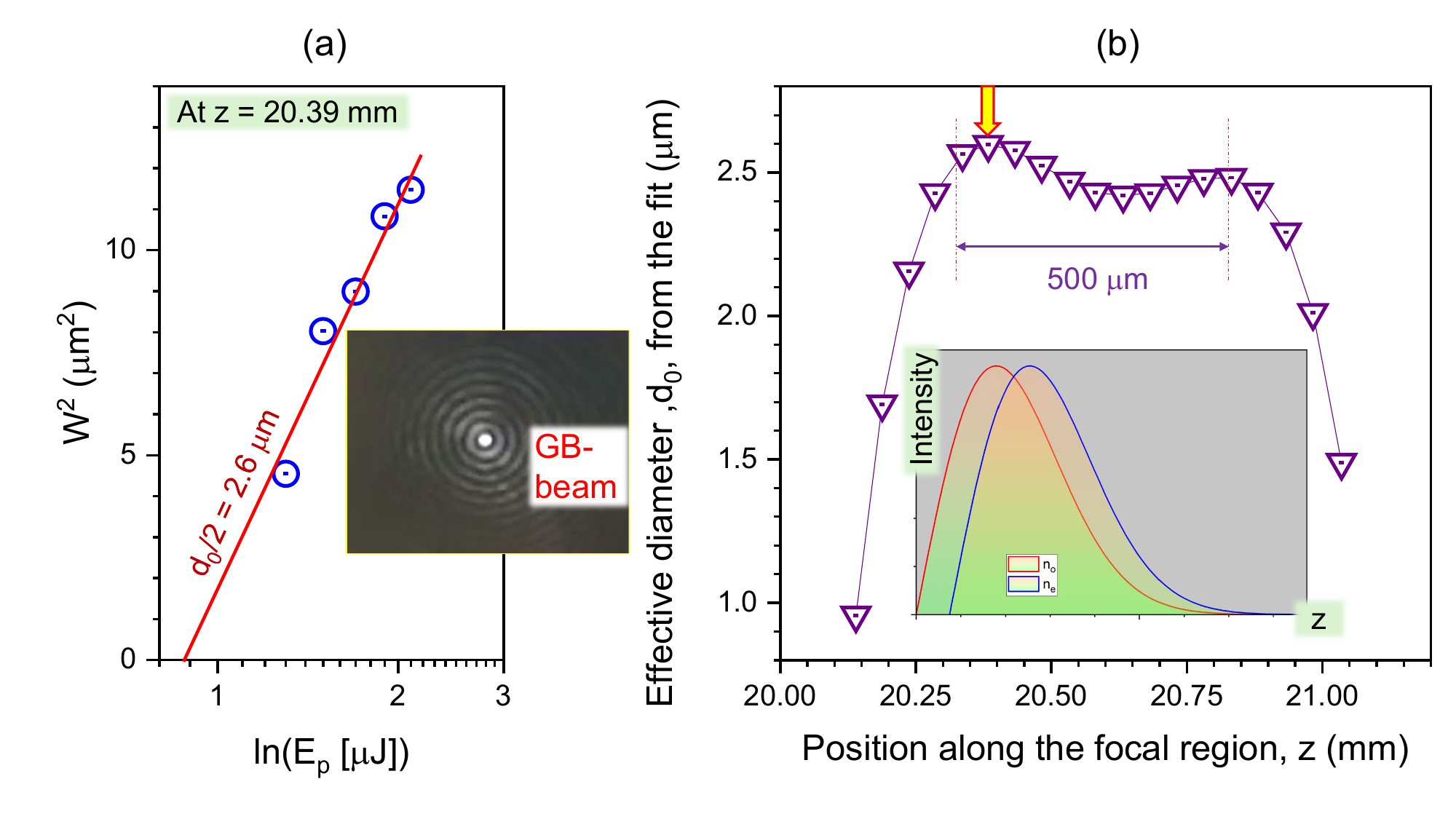}
\caption{\label{f-BLiu} Determination of the central core diameter of the GB-beam from the width (diameter), $W$, of ablation hole in 40-nm-thick \ce{AlOx} mask on Si. (a) Diameter ablated in alumina mask $W^2(\ln(E_p))$. The $slope$ of the best fit defines the radius $d_0/2 = \sqrt{slope/2}$ of the central core. The inset photo shows an image of the GB-beam at the alignment position (see Fig.~\ref{f-fab}(a)). (b) Determination of the effective diameter $d_0$ along the focal region of the GB-beam. A sample of b-Si was placed at the position marked by an arrow, which corresponded to $d_0 = 5.2~\mu$m. The axial extent of the non-diffracting region was $\sim 500~\mu$m. Circularly polarised light was used, which caused doubling of the characteristic axial GB-beam profile (Eqn.~\ref{e-ax}) shown in the inset for ordinary and extraordinary refractive indices of the form birefringent structure of diffractive axicon $n_{o,e}$.   
}
\end{figure}

\subsection{Geometrical parameters of the focal region produced by an axicon}
Axicon is a conical lens that transforms an incoming plane wave of a Gaussian intensity distribution into a Gauss-Bessel beam with a long axial extension. Refractive and diffractive axicons are used. For the refractive axicon, the cone acts as a prism sending the incoming light into a tilted beam, making a half-cone angle $\gamma$ (with optical axis). Refractive axicons are defined by the wedge angle required to form the full $180^\circ$ angle at the tip of the axicon (same angle at the base of the axicon), 
Then, from Snell's law $\sin(\alpha +\gamma) = n_{ax}\sin\alpha$, where the refractive index of axicon $n_{ax} = 1.4$ (for silica). 

The diffractive axicons are defined by the ring angle (peak-to-peak angle), which is equivalent to the full-cone angle $2\gamma$ of the refractive axicon; we used $\gamma \sim 1^\circ$ axicon (the ring angle 2$^\circ$).  

Length of the central non-diffracting part of the Gauss-Bessel beam is defined by the diameter $D$ of the beam as $Z_{max} = \frac{D}{2\sin\gamma} = \frac{2}{\pi}NZ_{RB}$, where the number of rings $N = D\sin\gamma/\lambda$ and $Z_{RB} = \frac{\pi^2 k}{2k_\perp^2}$ is the Rayleigh length associated with the asymptotic width of an individual ring ($Z_{RB} = 2.656$~mm for $\lambda = 1030$~nm and $\gamma = 1^\circ$). The component of wavevector perpendicular to propagation $k_\perp = k\sin\gamma$, where the wavevector $k = 2\pi/\lambda$. The GB-beam central spot size is typically defined as twice the first zero of the Bessel function $J_0(k_\perp\rho)$, where $\rho$ is the radius
$d_0 = 2\rho_0= \frac{4.816}{k_\perp} = \frac{4.816}{k\sin\gamma}$ ($45.2~\mu$m for $\lambda = 1030$~nm and $\gamma = 1^\circ$); the full width half maximum (FWHM) central cross section is $d_0^{\mathrm{FWHM}} = 0.36\frac{\lambda}{\sin\gamma}$ (21.2~$\mu$m)~\cite{Franc}. 

For actual fabrication, these values of $d_0$ and $Z_{max}$ are further down-sized using a 4f relay telescope; here we used $M_d = 10^\times$ demagnification which also correspondingly increased cone angle $\gamma'$ (in air for the sample position) $\sin\gamma = \sin\gamma'/M_d$  as described in the next section, Sec.~\ref{Sec-laser}, and shown in Fig.~\ref{f-fab}.

The energy flux associated with any ring of the Gauss-Bessel beam is equal to that of any other ring and of the central spot. The optimum central
spot illumination efficiency is given by~\cite{herman1991production}:
\begin{equation}\label{eq:efficiency}
    \epsilon_{op} = \frac{P_{ring}}{P_{total}} = \frac{2}{\sqrt{e}}\times \frac{\lambda}{w_0\sin\gamma} \equiv \frac{2}{\sqrt{e}}\times \frac{1}{N}
\end{equation}
\noindent where $w_0$ is the $1/e^2$-radius of the Gaussian beam intensity, i.e., the diameter of the laser beam on the axicon is $D = 2w_0 = 3.5$~mm in this study. Then, the length of non-diffractive region $Z_{max} = 100~\mu$m, the number of rings $N = 59$, and the efficiency of central spot illumination $\epsilon_{op} = 2\%$; for comparison, the G-beam efficiency of light collection into the focus is 100\%. 

The axial intensity distribution along the GB-beam propagation (z-axis) is given~\cite{Franc}:
\begin{equation}\label{e-ax}
    I_{GB}(z) = P_0\frac{8\pi z}{\lambda}\frac{\sin^2\gamma}{w_0^2}\exp\left[ -2\left(\frac{z\sin\gamma}{w_0}\right)^2\right],
\end{equation}
\noindent where $w_0$ is the waist of the Gaussian beam incident on the axicon and $P_0$~[W] is the peak power of the Gaussian beam. 

The axial intensity (Eqn.~\ref{e-ax}) maximum for the diffractive axicon has two axially-shifted positions for two perpendicular polarisations. This results in two distinct peaks when a circularly polarised incident beam is used~\cite{21jp024002}. It is caused by the form birefringence with two distinct refractive indices. This was the case for our study with circularly polarised beam ($\lambda/4$-plate was inserted into the laser output). Hence, the final $10^\times$ down-sized GB-beam had double length of $\sim 500~\mu$m; for one polarisation, $Z_{max}/10\approx 266~\mu$m is expected for the used setup/geometry (see Sec.~\ref{Sec-laser} and Fig.~\ref{f-BLiu}).

\subsection{Laser micro-machining setup} \label{Sec-laser}
For the GB-beam we used Workshop of Photonics (WOP) fs-fab station with a Carbide (Light Conversion) laser at 40~W, fundamental harmonic 1030~nm, repetition rate 602.7~kHz, pulse duration 201~fs (Fig.~\ref{f-fab}). A diffractive axicon with a ring angle of $2\gamma= 2^\circ$ was used (Holo OR). A combination of $f = 200$~mm lens (placed close to the axicon) with an objective lens MPlan APO NIR 10$^\times$ Mitutoyo, $NA = 0.26$ and $f = 20$~mm was used to form a 4f system for the imaging and downscaling 10 times the intensity distribution on the sample (Fig.~\ref{f-fab}(b)). 
Observation and inspection were carried out by MPlanApoNIR $2^\times$ $NA = 0.055$.

In a separate experiment, determination of the ablation threshold at the central core of the GB-beam was carried out using a 40-nm-thick \ce{AlOx} mask on Si wafer surface and applying the standard Liu method developed for G-beams~\cite{Liu82}. Ablation of \ce{AlOx} mask makes a very distinct hole and is useful to determine the central focal spot of GB-beam at the fabrication position; such masks were used in the fabrication of light trapping surface of Si solar cell~\cite{24aem2400711}. The diameter of the ablated hole in the nano-thin alumina mask vs fs-pulse energy was plotted $W^2(\ln[E_p])$ for different positions along the non-diffracting zone using the same objective lenses and setup described above (Fig.~\ref{f-BLiu}(a)). The sample position for fs-laser machining was decided by axial location, where the steepest dependence of the diameter of the ablated hole on the pulse energy was observed (Fig.~\ref{f-BLiu}(a)). The corresponding threshold pulse energy was $E_p^{(th)} \approx 900$~nJ (entire pulse) and the effective beam waist for the ablation at the center core was $d_0/2 = 2.6~\mu$m. The estimate of an average fluence per central core (per pulse) for the used efficiency $\epsilon_{op} = 1\%$ of GB-beam (1\% is considered instead of 2\% due to circularly polarised beam and separation of the intensity maxima into two peaks for two perpendicular polarisations). One finds $F_p^{(th)} = \frac{\epsilon_{op}E_p^{(th)}}{\pi(d_0/2)^2} = 0.043$~J/cm$^2$. 
The difference of diameters at the first-zero and FWHM of the GB-beam for the used geometry is $d_0/d_0^{\mathrm{FWHM}} = 2.13$. Then the energy in the central core per $d_0^{\mathrm{FWHM}}$ for the ablation threshold corresponds to the threshold fluence $F_p^{(th)}\times 2.13^2 = 0.2$~J/cm$^2$, which exactly matches that for Si. The $d_0^{\mathrm{FWHM}} = d_0/2.13 = 5.2/2.13 = 2.44~\mu$m. 

The analysis of focusing using geometry outlined above for diffractive axicon $\gamma = 1^\circ$ and $10^\times$ demagnification, corresponds to the center core energy of 1\% of that incident, which is focused onto a spot of $d_0^{\mathrm{FWHM}} =  2.44~\mu$m. These conditions will be used for analysis of fs-laser modification (melting and ablation) of b-Si and determination of the fluence at the center core (defined by $d_0^{\mathrm{FWHM}}$).

\section{Results}
Ablation and energy deposition onto b-Si was investigated using 1030~nm/200~fs pulses at different spacing between pulses in a linear scan. Gaussian and Gauss-Bessel fs-laser beams were used with comparable size of the central focal spot: 2.8~$\mu$m (Gaussian) and 2.6~$\mu$m (center-core Gauss-Bessel) diameters. 

\begin{figure}[tb]
\centering\includegraphics[width=0.8\textwidth]{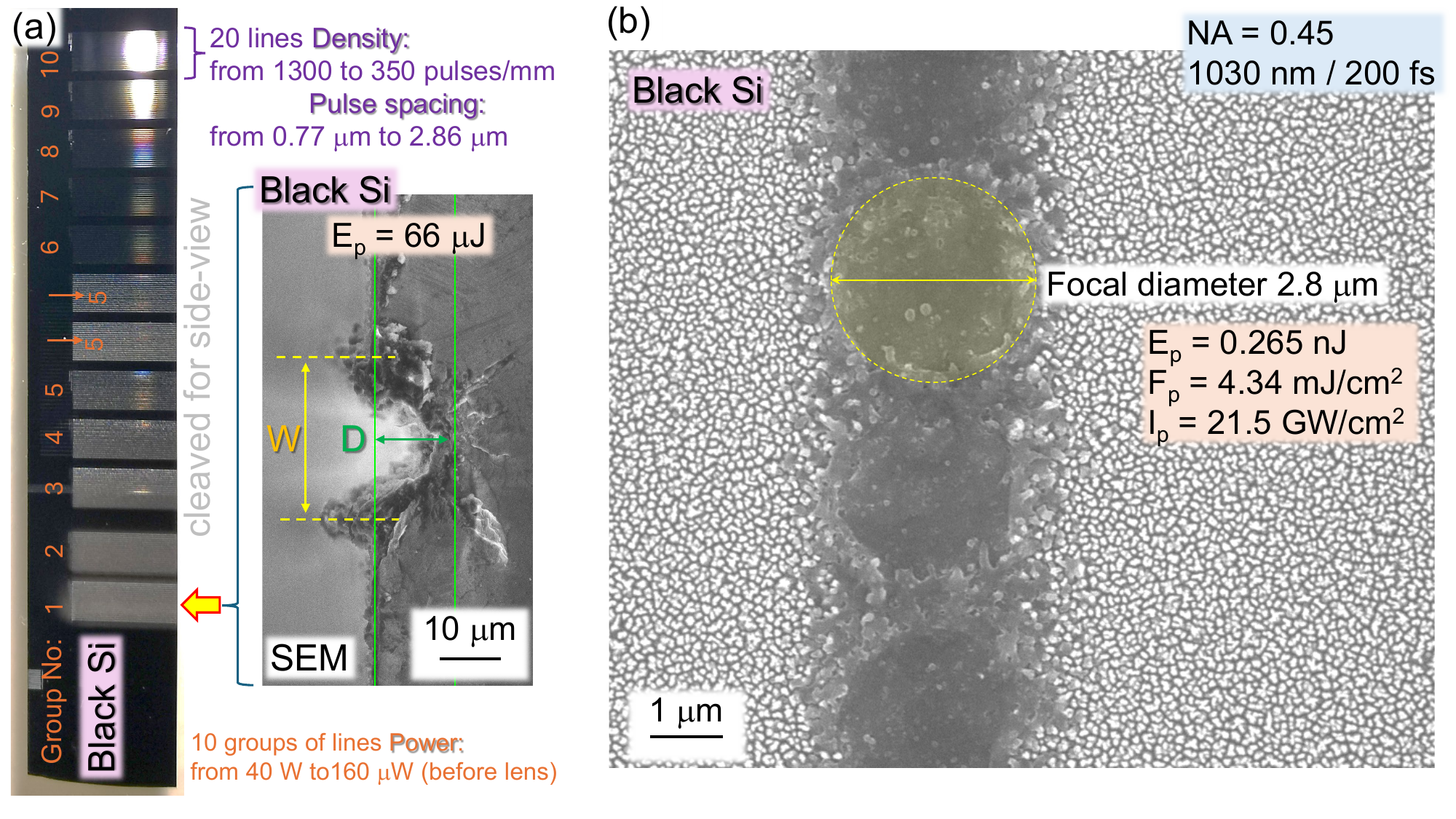}
\caption{\label{f-Gauss} Black-Si ablation by Gaussian fs-laser beam. (a) Photo of fs-laser machined regions and side view SEM image of ablation crater. (b) SEM image of laser-ablated sites with distinct imprints of 2.8~$\mu$m focal spots on the surface of b-Si. The diffraction limit $1.22\lambda/NA = 2.8~\mu$m for $\lambda = 1030$~nm wavelength and objective lens of numerical aperture $NA = 0.45$.  }
\end{figure}
\begin{figure}[tb]
\centering\includegraphics[width=0.8\textwidth]{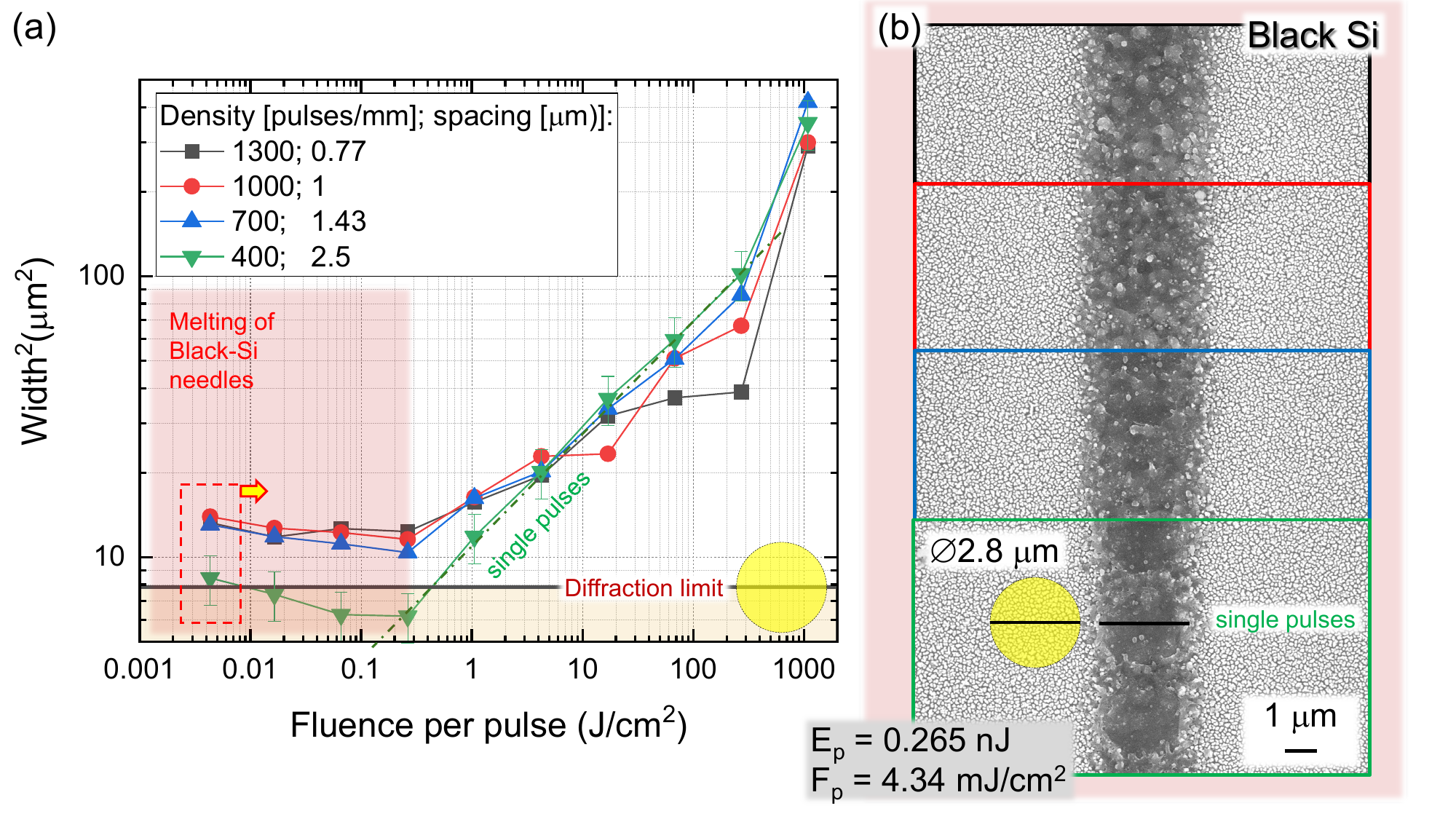}
\caption{\label{f-widG} (a) The width vs. single pulse fluence $W^2(F_p)$ for different pulse densities (pulse-to-pulse separation/spacing) ranging from just separated pulses to strongly overlapped. Error bars for single pulse data are $20\%$. (b) SEM images of the lowest fluence modification made by remelting the b-Si surface.    }
\end{figure}
\begin{figure}[tb]
\centering\includegraphics[width=0.8\textwidth]{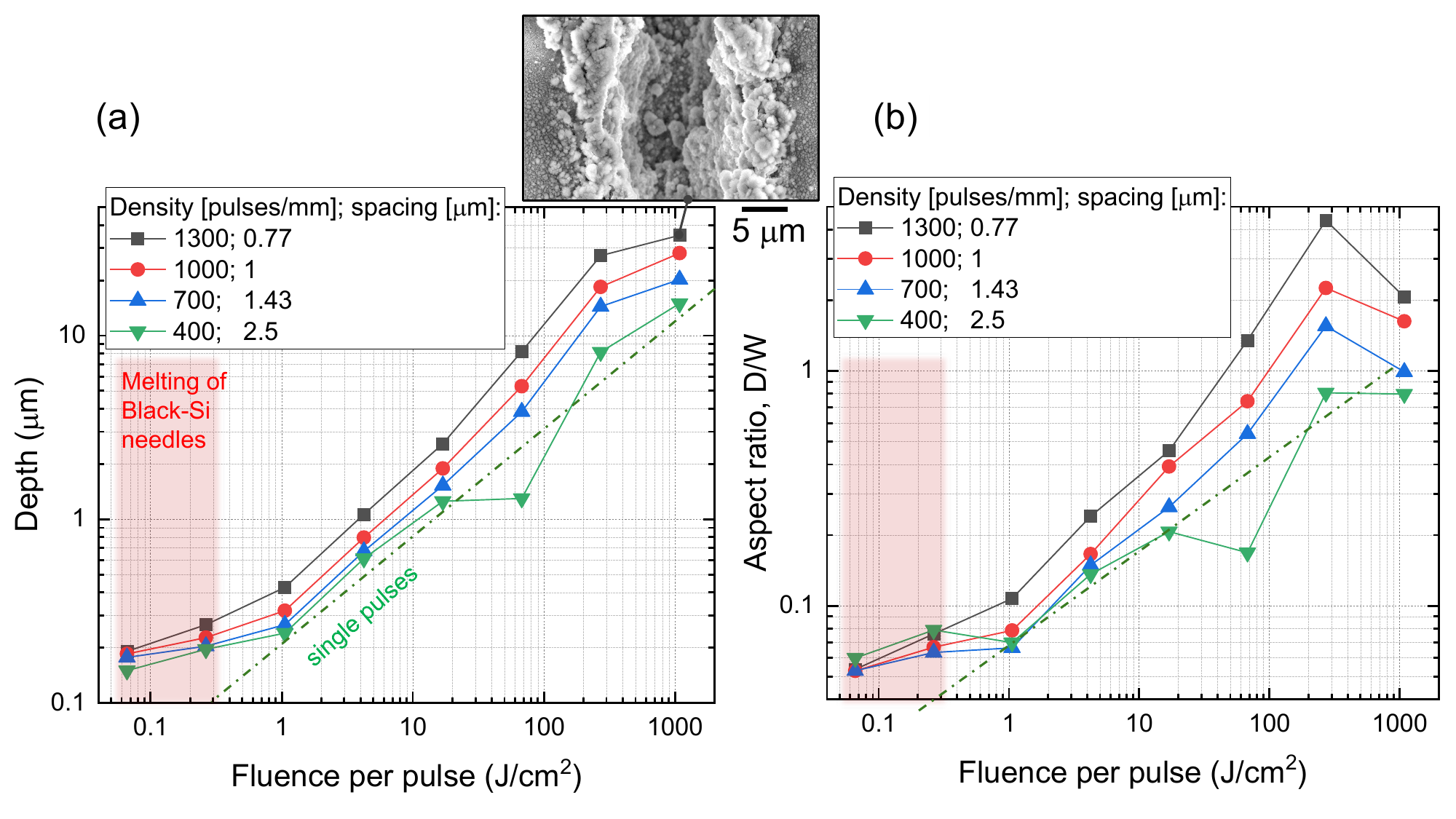}
\caption{\label{f-ratio} (a) Depth vs. fluence per pulse.  The top inset shows a SEM image of high-fluence, small spacing ablation. (b) The aspect ratio vs. fluence per pulse.    }
\end{figure}

\subsection{Gaussian beam}
The diameter of the Gaussian (G-)beam can be estimated as $d_G = 1.22\lambda/NA$ for the objective lens of numerical aperture; $NA = 0.45$ in this study. It is the diameter of the Airy disk, i.e., between two minima (zero intensity) in cross section for a plane wave focusing limited only by diffraction. In terms of energy under the envelope of the Gaussian beam intensity profile (86.5\%) and within the Airy-disk for the plane wave (84\%), both are very similar.  

Figure~\ref{f-Gauss} shows a typical sample with lines laser ablated at different pulse energies and spacing between adjacent irradiation spots. A sample of b-Si was cleaved across the line pattern for the side view observation of the ablation depth $D$ and width $W$ (Fig.~\ref{f-Gauss}(a)). At very low pulse fluence, well below the ablation threshold of Si 0.2~J/cm$^2$~\cite{Bonse}, surface modification of b-Si had clear imprint of melted nano-needles with discernable rim with diameter of 2.8~$\mu$m, which matched the focal spot size for the used $NA = 0.45$ objective lens $2r_G = 1.22\lambda/NA = 2.8~\mu$m. This diameter was used for the determination of the average pulse fluence $F_p = E_p/(\pi r_G)^2$.

The dependence of ablation/modification width on the pulse fluence is presented as $W^2(\ln[F_p])$ in Fig.~\ref{f-widG}(a). Log-log presentation is chosen to better reveal the change of slope, which is linked to the mechanism of the modification. This is similar to the Lin-log presentation of the same dependence known as the Liu method~\cite{Liu82}. For the largest separation between pulses close to the focal diameter, the threshold of modification is close to 0.2~J/cm$^2$, which is the expected threshold for Si ablation. However, at much lower fluence $\sim 4$~mJ/cm$^2$ by a factor of 50, a clear surface modification was observed. It is linked to the melting of nano-needles of b-Si and was lower as observed on flat Si~\cite{Jan}. There was almost no change in the width of melt-modification. This confirms that direct energy deposition occurred only over the focal spot and the lateral heat diffusional spread was negligible, as shown in Fig.~\ref{f-widG}(b). Interestingly, at the maximal used fluence, the width was increasing at a steeper rate vs $F_p$, departing from the expected linear dependence of the Liu plot. This could be linked to the better energy deposition at the surface of b-Si due to low reflectance. Figure~\ref{f-ratio} shows the evolution of the depth and aspect ratio of the ablated groove. For the low pulse overlap, close to single pulse ablation conditions, the threshold fluence projects to the expected 0.2~J/cm$^2$ for Si.

\subsection{Gauss-Bessel beam}  

\begin{figure}[tb]
\centering\includegraphics[width=0.9\textwidth]{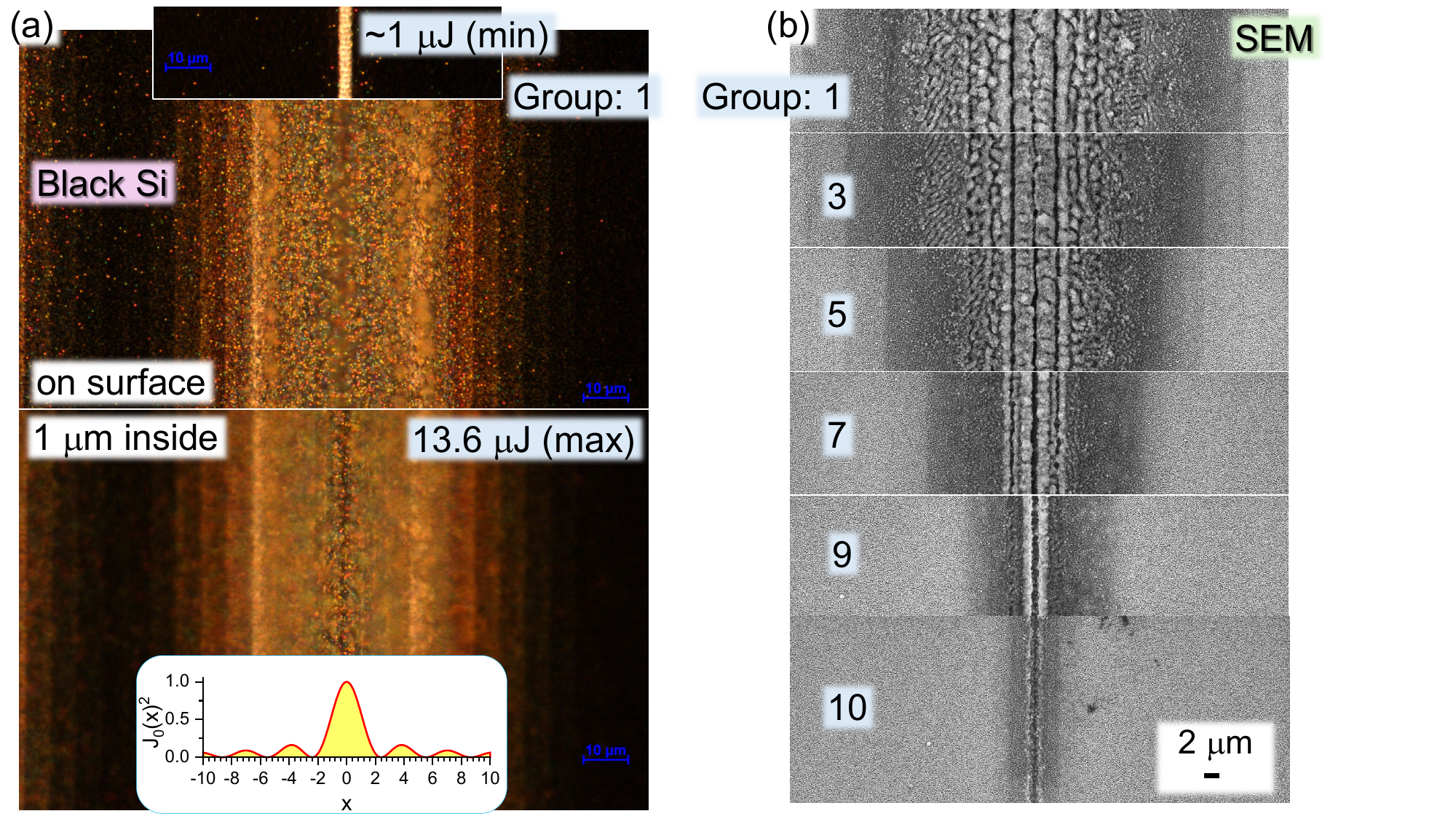}
\caption{\label{f-Bies} Black-Si ablation by Gauss-Bessel beam.  (a) Dark-field image (with $NA = 0.9$ objective lens) of ablated line at high pulse energy $E_p$. The top inset shows the lowest pulse energy ablated line and the bottom inset shows the Bessel function where $x = k_\perp\rho$. (b) SEM top-view images of ablated lines at different pulse energies (by group number). Ripples at off-center locations of GB-beam show that polarisation was linear (or elliptical) rather than circular, which is aimed at the center core.    }
\end{figure}

To calculate the energy, fluence and irradiance (intensity) per pulse at the center core of the Gauss-Bessel (GB-)beam is not trivial, as in the case of the Gaussian beam, since the energy of the GB-beam is spread over the number of rings, which carry the same energy and converge onto optical axis at different locations along the propagation. Moreover, for the diffractive axicon, there are two axially shifted intensity distributions (Eqn.~\ref{e-ax}) for two perpendicular polarisations~\cite{21jp024002}. This reduce energy distribution per ring twice. It is noteworthy that the Liu method can be used for different non-Gaussian beam shapes~\cite{NotG}. 

\FloatBarrier
\FloatBarrier
\raggedbottom
We adopted the established Liu method~\cite{Liu82} for G-beam to measure the effective diameter of the focal spot at different locations along the focal region (non-diffractive zone of GB-beam) as described in Sec.~\ref{Sec-laser}. The test sample for measurement of ablated diameter vs the pulse energy was 40-nm-thick \ce{AlOx} coated mask on Si. It was placed at different positions of $\sim 1$~mm long focal region of the GB-beam using the same axicon and focusing conditions.  

According to the Liu method~\cite{Liu82} for the G-beam, the diameters, $W$, of the abated pits (or width of the ablated line) on the sample at different pulse energies are determined. The intercept of a linear fit of $W^2[\ln(E_p)]$ corresponds to a threshold pulse energy $E_{th}$ (also threshold fluence $F_{th}=E_{th}/(\pi w_0^2)$) while the $slope$ corresponds to $2w_0^2$:
\begin{equation}\label{eG}
 W^2 = 2w_0^2\ln(E_p/E_{th}).   
\end{equation}
\noindent The waist (radius) of the Gaussian beam at the focus is $w_0 =\sqrt{slope/2}$ (Eqn.~\ref{eG}). 

We verified this experimental procedure of focal diameter determination using ablation of \ce{AlOx} mask on Si using GB-beam for the center-core ablation/modification as described in Sec.~\ref{Sec-laser} and shown in Fig.~\ref{f-BLiu}. It was determined that the ablation threshold of Si 0.2~J/cm$^2$ corresponds to the energy at the center core over its FWHM cross section. Also, as in the case of G-beam, the diameter of the center core (energy deposition) is determined from the low pulse energies. 

Typical patterns of GB-beam ablated b-Si are shown in Fig.~\ref{f-Bies}. The dark-field optical image in (a) shows a pattern at high energy/fluence at the surface of the sample and 1~$\mu$m depth. SEM images at the same high fluence and down to low energy $E_p$ when only the center core is discernible are overlayed in (b). Counterintuitively, the width of the central opening was narrower for the high $E_p$. This is caused by the formation of a high-aspect-ratio central groove and redeposition/oxidation around the center core.  For this particular reason, analysis by Eqn.~\ref{eG} cannot be applied for the b-Si; however, it was verified for the mask hole ablation of alumina-coated Si as described previously (Sec.~\ref{Sec-laser}).
\begin{figure}[tb]
\centering\includegraphics[width=0.75\textwidth]{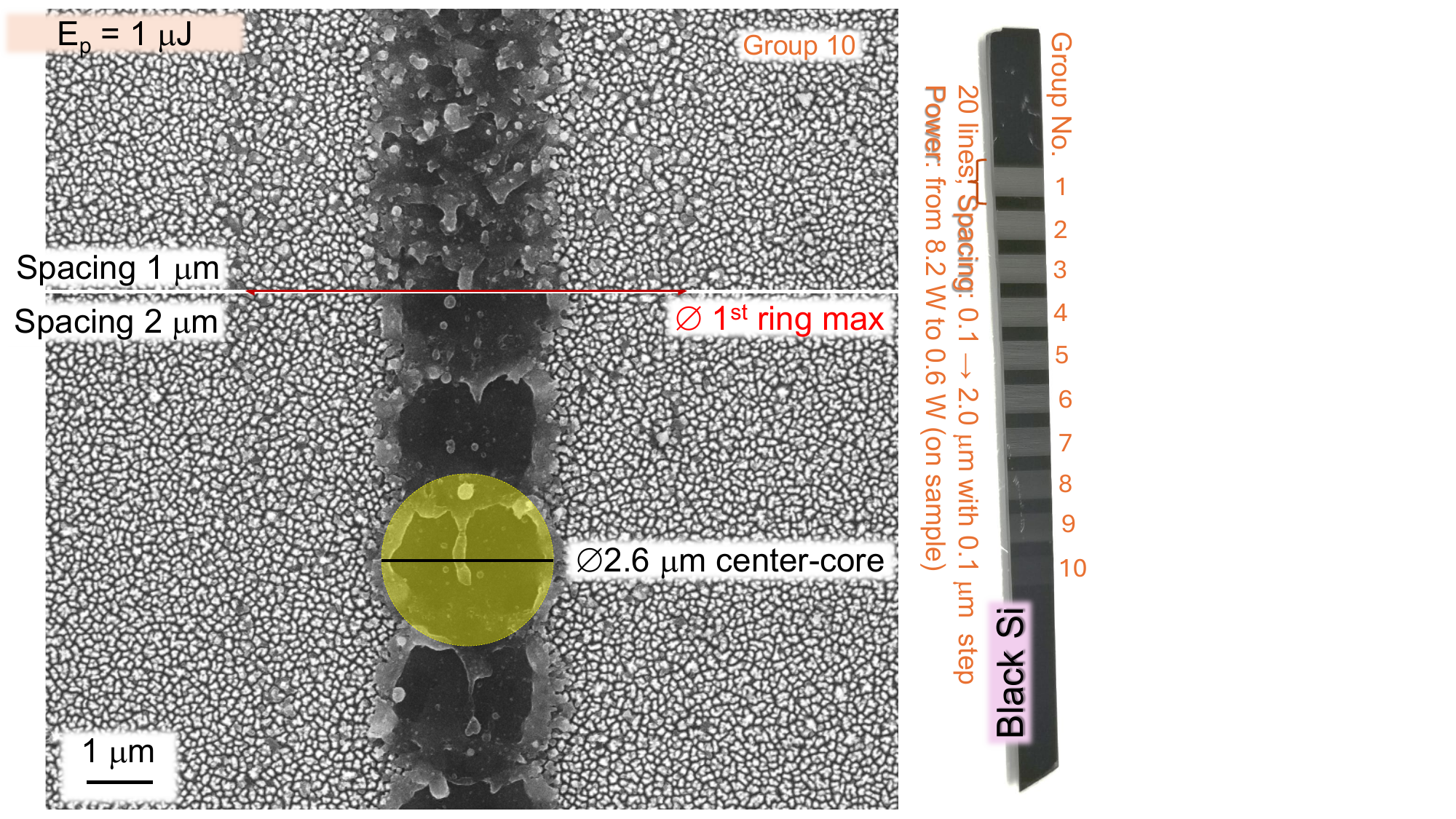}
\caption{\label{f-Bmin} SEM image of b-Si ablated by Gauss-Bessel beam at lowest pulse energy. It was used to determine the diameter of the center core. The right-side inset shows a photo of the b-Si sample cleaved across all the patterns recorded under different conditions for the side-view imaging of the depth and width. The position of the first ring maximum is barely recognisable (marked by a red arrow).   }
\end{figure}

At the lowest pulse energies, a clear molten b-Si surface is observed with clearly discernible 2.6~$\mu$m diameter (Fig.~\ref{f-Bmin}). It is very close to the expected FWHM diameter of 2.44~$\mu$m as determined from the geometrical conditions and focusing (Sec.~\ref{Sec-laser}). This diameter was used to calculate pulse fluence at the center core as previously validated in the adaptation of the Liu method. The sample position for the alumina mask ablation on Si and b-Si was the same as all the beam delivery optics. 

\begin{figure}[tb]
\centering\includegraphics[width=0.8\textwidth]{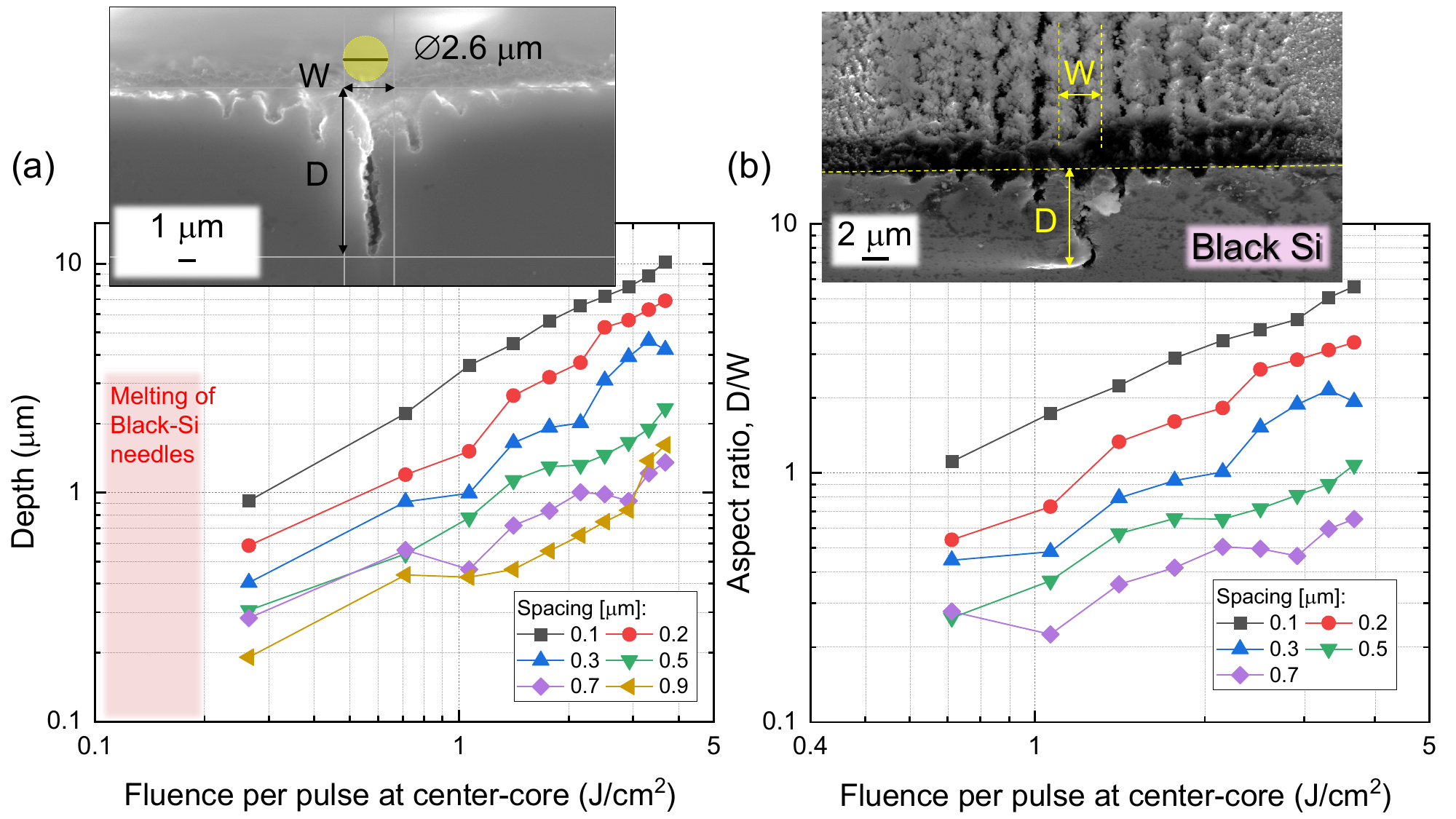}
\caption{\label{f-GB} The depth $D$ (a) and aspect ratio $W/D$ vs effective fluence per center-core $d_0 = 2.6~\mu$m diameter. The fluence was calculated as $E_p$ over the area of diameter $d_0$ (experimentally observed at lowest fluences); a factor of 1\% was used for energy redistribution into the center core as determined in Sec.~\ref{Sec-laser}. 
}
\end{figure}

Figure~\ref{f-GB} shows the depth and aspect ratio evolution with fluence per center core at different pulse-to-pulse spacings. Close to linear dependence was observed over the span of one order of magnitude in $F_p$ (per pulse per center core). Since the large surface area is affected, the determination of width $W$ had a large uncertainty (obvious also from Fig.~\ref{f-Bies}(b)). This is why some data points are excluded from the aspect ratio plot in (b). A formidable high aspect ratio $>5$ was observed for the center fluence $> 2.5$~J/cm$^2$. A deep central ablation groove was formed under those conditions, and the imprint of up to 3 rings was present on the surface. Apparently, the energy deposition has a larger aspect ratio as compared with the G-beam case. This can be facilitated by the anti-reflective property of b-Si, deeper optical light penetration due to 1030~nm wavelength being close to the bandgap of Si. 

\section{Discussion}

It is shown here that the Liu methodology of determination of ablation threshold, originally demonstrated for G-beams, is applicable to the center core GB-beams. When the FWHM of the central core was taken as the diameter of the energy deposition, the efficiency of energy at the center beam was calculated using ideal Bessel beam properties at the employed focusing conditions. This is useful for comparison of G- and GB-beam laser machining, at least at the semi-quantitative level. Exact Bessel-like beam geometry and properties are affected by focusing, filling/clipping at the entrance apertures, and demagnification telescope alignment in addition to the type of optical element, refractive, diffractive, metasurface, and the quality, which sets phase imperfections. 

Anti-reflective property of b-Si allowed very precise control of energy deposition with clear accumulation effect revealed in Fig.~\ref{f-widG}: the smaller spacing, i.e., the larger pulse overlap, the slope of $W^2$ vs $F_p$ dependence is less steep, especially towards smaller fluencies. Apparently, the cumulative effect of irradiation is enhancing energy deposition and remelting of nano-needles. Larger droplets at stronger overlap signify larger surface temperatures over larger volumes, which was able to minimize surface area by forming spheroidal droplets. Such controlled melting 
has potential for surface doping of materials when b-Si is coated by other elements which are potential dopants for n- or p-type conductivity as well as formation of intra-band for IR sensors~\cite{hydo} 
Black-Si produced by plasma etch has a surface with native oxide similar to the flat Si with detectable F presence~\cite{16semsc221}. This makes doping and remelting of nano-needles with surface deposited dopants a promising avenue for tailoring surface properties and chemistry.  

We can put forward a conjecture that light localisation between nano-needles and at the interface Si-air with large E-field components normal-to-the-interface contributes to energy deposition. Inducing defects which enhance absorption of subsequent pulses drived energy deposition onto this antireflective surface.  Indeed, it corresponds to the p-polarisation rather than s-polarisation. and has lower reflectance (see simulations in ref.~\cite{20n873}).  

\subsection{Ablation threshold of black-Si}

Easily recognisable surface modification is observed at a much lower threshold than expected for Si ablation at 0.2~J/cm$^2$ per pulse (for optically flat Si)~\cite{Bonse}. Ablation thresholds of metals (m) and dielectrics/semiconductors (d) is related to ionisation and are estimated from the energy deposition which exceeds the cumulative binding energy (enthalpy of vaporisation) $\epsilon_b$ and electron work function (for metals) $w_e$ or $\epsilon_b$ and ionisation potential  $J_i$ for dielectrics and semiconductors~\cite{Gamaly2002}:  
\begin{equation}\label{Gam}
F^{(m)}_{th} = \frac{3}{8}(\epsilon_b + w_e)\frac{\lambda n_e}{2\pi},~~~~\\
F^{(d)}_{th} = \frac{3}{4}(\epsilon_b + J_i)\frac{l_s^*n_e}{A}.
\end{equation} 
where 
$l_s^* = c/(\omega\kappa)\equiv\lambda/(2\pi\kappa)$ is the absorption depth (the skin depth for the $E$ field) in the plasma with electron density $n_e$ and refractive index $n^* = n +i\kappa$ with $c$ and $\omega$ being the speed and cyclic frequency of light, respectively; $A$ is the absorption coefficient (for good metals $A\approx\frac{2\omega l_s^*}{c}$). At high intensity, the pulse can exceed the ionization threshold and the first ionization is completed (before the end of the pulse), at which the number density of free electrons saturates at the level $n_e \approx n_a$, where $n_a$ is the number density of atoms
in the target material. 

For ablation of dielectrics and semiconductors, which are ionized during the fs-pulse, a metal-like reflective plasma has $A\approx 0.5$; i.e., half of the light is absorbed by the strongly excited material. The exact absorbed portion $A = 1-R$, where $R$ accounts for the reflected portion of the laser pulse, can be precisely calculated from the refractive index $n^*\equiv n +i\kappa$ as $A = 4n/[(n + 1)^2 + \kappa^2]$. 
The ablation threshold of metals (Eqn.~\eqref{Gam}) scales with wavelength $F^{(m)}_{th} \propto \lambda$~\cite{Gamaly2002}.
Equation~\eqref{Gam} is based on the required energy budget to evaporate material $\epsilon_b$ and to ionize it: $w_e$ for metals and $J_i$ for dielectrics. These thresholds are confirmed by experiments for metals and dielectrics~\cite{Gamaly2002}.  

For unperturbed Si, $\epsilon_b = 383$~kJ/mol, the first ionisation potential $J_i = 786.52$~kJ/mol (8.152~eV per atom) and typical $n^* = 3.5 +i0.01$ corresponding to a low doping Si $n_e = 1\times 10^{16}$~cm$^{-3}$ (or resistivity $\sim 1~\Omega.$cm); the skin depth reciprocal to the absorption coefficient (for intensity $E^2$) $l_s=1/\alpha = (4\pi\kappa/\lambda)^{-1} = 8.2~\mu$m; $\alpha = 1.22\times 10^3$~cm$^{-1}$, $A = 69.1\%$ at $\lambda = 1030$~nm wavelength. 
Upon fs-laser irradiation, the electron density is approaching the critical plasma density $n_{cr} = \omega^2\varepsilon_0m_e^*/e^2$ at which the real part of permittivity is decreasing and the imaginary part is increasing $\varepsilon^* = \varepsilon_1 +i\varepsilon_2$. Under such conditions, the most efficient energy deposition is taking place (energy per volume), and dielectric breakdown follows. For $\lambda = 1030$~nm, $n_{cr} = 1.05\times 10^{21}$~cm$^{-3}$, where $\varepsilon_0$ is the permittivity of free space, $m_e^*$ is the effective mass of electron. The dielectric breakdown is defined by the real part of permittivity becoming zero:  $\varepsilon_1 = (n^2 -\kappa^2) \equiv 0$ and $n=\kappa$; $\varepsilon_2 =  2n\kappa$ defines absorption losses of light in the sample. Assuming that photo-excited Si approach $n = \kappa = 0.2$, one finds $A = 54.1\%$ and for the plasma density close to critical $n_e \approx n_{cr}$, at which the ablation threshold of Si becomes $F_{th}^{\mathrm{Si}} = 0.232$~J/cm$^2$  (Eqn.~\ref{Gam}); for these conditions $l_s\equiv l_s^*/2 = 410$~nm (close to the height of b-Si nano-needles~\cite{25srm316}) $\alpha = 2.44\times 10^4$~cm$^{-1}$. The threshold is close to the experimentally observed ablation threshold on flat Si by ultra-short laser pulses 0.2~J/cm$^2$~\cite{Bonse}. 

When Si is turned into black-Si, the material properties $\epsilon_b$ and $J_i$ remain the same, however, the absorption coefficient $A$ is effectively increased due to the reduced reflectivity. Indeed, when there is no transmission $T = 0$ (an optically thick sample), the energy conservation $A+R+T = 1$ demands $A = 1-R$. This can be interpreted as a reduced ablation threshold expected for the anti-reflective surface and observed experimentally in this study (Fig.~\ref{f-widG}). 

Ablation (removal of material) takes place when the pool of energy absorbed by electrons transfers it to the lattice and heats it above the evaporation conditions (the binding energy). The temperature of electrons in the skin depth is $T_e = (1-R)F_p/(l_sc_en_a)$~\cite{Gamaly2002}, where $F_p$ is the fluence per pulse, $c_e\sim 3/2$ is the electron specific heat, which acquires the value of 3/2 of the ideal gas after full ionization of material, and $n_a$ is the atomic number density of the material (for Si $n_0 = \rho_{Si} N_{av}/M_{Si} = 4.99\times 10^{22}$~cm$^{-3}$, here the mass density $\rho_{Si} = 2.33$~g/cm$^3$, molar mass $M_{Si} = 28.08$~g and $N_{av}$ is the Avogadro number). This energy 
deposited into electrons acts as an energy reservoir for ablation after its transfer to lattice atoms/ions, with $T_e$ reaching $\sim$2~keV in typical metal and silicon drilling~\cite{04apa1555}. The maximum of material removed by a single pulse can be estimated from energy conservation~\cite{Gamaly2002}: 
$V_{max} = (1-R)E_p/(n_aE_{at}/N_a)$,
where $E_{at}$~[J/mol] is the molar enthalpy of atomization and $N_A$ the Avogadro constant. Also here, the low reflectivity maximises material removal facilitated by larger absorbance (Eqn.~\ref{Gam}), which is additionally enhanced due to p-pol. at a large angle of incidence at the laser pulse and side-walls of nano-needles $\theta_i\sim\theta_B\approx 75^\circ$, where $\theta_B$ is the Brewster angle.

\section{Conclusions and Outlook}


It is demonstrated that the anti-reflective surface of b-Si facilitates very precise energy deposition at the focal spot with a smooth transition from the ablation (removal of material via ionisation above 0.2~J/cm$^2$) to a well-controlled energy deposition and melting of the surface nanotexture at fluencies 4~mJ/cm$^2$. The width of the molten - re-solidified region closely follows the focal spot size. The reduction of ablation threshold by $50^\times$ is related to the increased absorbance (reduced reflectance) of b-Si. Change of reflectance of Si vs b-Si $\frac{R}{R_b} = \frac{1-R_b}{1-R}\equiv\frac{A_b}{A}\approx \frac{40\%}{1\%} \sim 40$. This factor of ablation threshold reduction is achievable due to optical properties and energy deposition into the very same Si (same binding energy and ionisation potential), governed by geometrical factors of the surface texture. It is noteworthy that an estimation of such a reduction factor of ablation threshold is only qualitative since the energy deposition is a dynamic fast process governed via free carrier (electron-hole) generation and self-action on the incoming light, which determines the energy deposition via the instantaneous complex permittivity $\varepsilon^*$ of irradiated surface of b-Si. Further experimental efforts to explore the reduction of ablation thresholds due to the texture of the same material as the host substrate could open new methods for controlling resolution and feature size of the nanoscale energy deposition.   

\small\begin{acknowledgments}
This study was carried out via ARC DP240103231 grant. This work was performed on the fs-fab laser station at Swinburne, which is jointly funded with the Melbourne Centre for Nanofabrication (MCN),  the Victorian Node of the Australian National Fabrication Facility (ANFF). 
We thank Workshop-of-Photonics (WOP) Ltd., Lithuania for the patent licence and technology transfer project by which the industrial fs-laser fabrication setup was acquired for Nanolab, Swinburne.
\end{acknowledgments}


\bibliography{aipsamp}

\end{document}